\documentclass[preprint,
12pt,showpacs,preprintnumbers,nofootinbib,nobibnotes,amsmath,amssymb,aps,prd]{revtex4-1}
\usepackage{epsfig}
\usepackage{graphicx}
\usepackage{bm}
\usepackage{times}
\usepackage{braket}
\usepackage{color}
\usepackage{slashed}
\usepackage{hyperref}
\usepackage{ulem}
\definecolor{Red}{rgb}{1.,0.,0.}

\definecolor{Blue}{rgb}{0.,0.,1.}

\definecolor{Green}{rgb}{0.,1.,0.}

\definecolor{Gray}{rgb}{0.5,0.5,0.5}

\definecolor{nicered}{rgb}{0.7,0.1,0.1}
\definecolor{nicegreen}{rgb}{0.1,0.5,0.1}
\hypersetup{colorlinks,citecolor=nicegreen,linkcolor=nicered}

\begin{document}

\newcommand{\beq}{\begin{eqnarray}}
\newcommand{\eeq}{\end{eqnarray}}
\newcommand{\ben}{\begin{enumerate}}
\newcommand{\een}{\end{enumerate}}
\newcommand{\non}{\nonumber\\ }
\newcommand{\jpsi}{J/\Psi}
\newcommand{\ppa}{\phi_\pi^{\rm A}}
\newcommand{\ppp}{\phi_\pi^{\rm P}}
\newcommand{\ppt}{\phi_\pi^{\rm T}}
\newcommand{\ov}{ \overline }
\newcommand{\zerot}{ {\textbf 0_{\rm T}} }
\newcommand{\kt}{k_{\rm T} }
\newcommand{\fb}{f_{\rm B} }
\newcommand{\fk}{f_{\rm K} }
\newcommand{\rk}{r_{\rm K} }
\newcommand{\mb}{m_{\rm B} }
\newcommand{\mw}{m_{\rm W} }
\newcommand{\im}{{\rm Im} }
\newcommand{\kks}{K^{(*)}}
\newcommand{\acp}{{\cal A}_{\rm CP}}
\newcommand{\pb}{\phi_{\rm B}}
\newcommand{\xeba}{\bar{x}_2}
\newcommand{\xsba}{\bar{x}_3}
\newcommand{\peas}{\phi^A}
\newcommand{\Dsl}{ D \hspace{-2truemm}/ }
\newcommand{\pvsl}{ p \hspace{-2.0truemm}/_{K^*} }
\newcommand{\esl}{ \epsilon \hspace{-2.1truemm}/ }
\newcommand{\psl}{ p \hspace{-2truemm}/ }
\newcommand{\ksl}{ k \hspace{-2.2truemm}/ }
\newcommand{\lsl}{ l \hspace{-2.2truemm}/ }
\newcommand{\nsl}{ n \hspace{-2.2truemm}/ }
\newcommand{\vsl}{ v \hspace{-2.2truemm}/ }
\newcommand{\zsl}{ z \hspace{-2.2truemm}/ }
\newcommand{\epsl}{\epsilon \hspace{-1.8truemm}/\,  }
\newcommand{\bfkk}{{\bf k} }
\newcommand{\calm}{ {\cal M} }
\newcommand{\calh}{ {\cal H} }
\newcommand{\calo}{ {\cal O} }

\def \appb{{\bf Acta. Phys. Polon. B }  }
\def \cpc{ {\bf Chin. Phys. C } }
\def \ctp{ {\bf Commun. Theor. Phys. } }
\def \epjc{{\bf Eur. Phys. J. C} }
\def \ijmpcs{{\bf Int. J. Mod. Phys. Conf. Ser.} }
\def \jhep{{\bf J. High Energy Phys. } }
\def \jpg{ {\bf J. Phys. G} }
\def \mpla{{\bf Mod. Phys. Lett. A } }
\def \npb{ {\bf Nucl. Phys. B} }
\def \plb{ {\bf Phys. Lett. B} }
\def \ppn{ {\bf Phys. Part. Nucl. } }
\def \ppnp{{\bf Prog.Part. Nucl. Phys.  } }
\def \pr{  {\bf Phys. Rep.} }
\def \prc{ {\bf Phys. Rev. C }}
\def \prd{ {\bf Phys. Rev. D} }
\def \prl{ {\bf Phys. Rev. Lett.}  }
\def \ptp{ {\bf Prog. Theor. Phys. }}
\def \zpc{ {\bf Z. Phys. C}  }
\def \jpg{ {\bf J.Phys.-G-}  }
\def \ap{ {\bf Ann. of Phys}  }

{\footnotesize

\title{Pion and Kaon form factors in the perturbative QCD approach}

\author{Shan Cheng} \email{scheng@hnu.edu.cn}

\affiliation{School of Physics and Electronics, Hunan University, 410082 Changsha, People's Republic of China} 

\date{\today}

\begin{abstract}
We present the most accurate calculation for the pion and kaon electromagnetic form factors in the framework of perturbative QCD,
where the power corrections up to twist-4 of the meson distribution amplitudes and
the next-to-leading-order QCD corrections up to subleading power are included.
In order to guarantee the gauge invariance of the meson to vacuum matrix element,
we take into account both assignments with the lowest Fock state and the high Fock state with an additional valence gluon.
Our results confirm the power behaviour of the twist expansion and
show  the chiral enhancement effect at subleading power in the PQCD approach.
We also estimate the $\mathrm{SU(3)}$ asymmetry for the kaon and pion form factors  and find that it is  smaller than $30 \%$.
\end{abstract}

\maketitle

\section{Introduction}

The Quantum chromodynamics (QCD) has two fundamental properties:   the quark confinement in the low energy region and the asymptotic freedom in the
high energy region. The confinement leads to the formation of the hadrons, while the asymptotic freedom of the strong interaction results in the
perturbative   QCD calculations.
When an energetic photon hit   a constituent parton (quark, antiquark or gluon, etc )  inside  a hadron,  one uses  a function " form factor"  to
describe the redistribution of the momenta of the parton  inside the hadron.
The form factor  therefore carries both the information of  hadron structure and  the hard scattering amplitude.
In order to calculate the form factor for a given transition process,
the factorization theory is developed to help one to separate the pertubative and nonperturbative contributions \cite{LepageZA,LepageFJ,EfremovQK}.
The electromagnetic (e.m.) form factor of pion, being the simplest but simultaneously the most fundamental QCD observed quantity,
attracts much attention  both in  theory \cite{GoussetYH,ff-pion-QCDSRs,ff-pion-LCSRs,ff-pion-PQCD,ff-pion-LQCD}
and in experiments  \cite{prd9-1229,prl95-261803,prl97-192001}.

The statements for the form factors  are rather different  in different theoretical approaches.
In the QCD factorization (QCDF) \cite{BenekeBR,BauerEW,Beneke:2002ph}, for example, the form factor is the nonpertuabtive input.
In the light-cone sum rules (LCSRs), one believes that the soft dynamics will provide the dominate contribution \cite{ff-pion-LCSRs}.
In the perturbative QCD (PQCD)  approach, however,  it is described by a hard scattering amplitude \cite{LiUN,ff-pion-PQCD} and can be
calculated perturbatively.
For the pion form factor, for instance, the lattice QCD (LQCD) evaluation is still available at a few points of the momentum transfer squared $Q^2$ so far
\cite{ff-pion-LQCD},  while the direct experiment measurements are credible below $3$ GeV$^2$ \cite{prl95-261803,prl97-192001}  too.
The LCSRs approach is reliable in the intermediate region $1 \leq  Q^2 \leq 15$ GeV$^2$ \cite{BelyaevZK},
and the prediction power of the  PQCD approach holds well in the large region $Q^2 \geq 10 \, \mathrm{GeV}^2$ with the inclusion of
the resummation effects.
In this paper we calculate  the higher power corrections to pion and kaon form factors up to twist-4 of the meson DAs,
with the aim to check the power expansion behaviour from one side, and from the other side to improve the theoretical accuracy in the framework of PQCD approach.

The rest of the paper is organized as follows.  In Sec.\ref{sec:power}, the PQCD calculation of the spacelike pion form factor is performed
by considering both the quark-antiquark and the quark-antiquark-gluon assignments.
In Sec.\ref{sec:PQCD}, we present the procedure of  the PQCD approach to calculate the pion form factor, several important issues are
highlighted.  Sec.\ref{sec:numerics} contains  the numerical results and we conclude in Sec.\ref{sec:conclusion}.

\section{Power corrections}\label{sec:power}

The pion form factor is defined by the nonlocal matrix element
\beq
\langle \pi^-(p_2) \vert J_\mu^{\mathrm{e.m.}} \vert \pi^-(p_1) \rangle \equiv e_q (p_1 + p_2) F_\pi(Q^2) \,,
\label{eq:ff-1}
\eeq
we are interest in the case that the smallness of relative distance is ensured by the "external reason",
says large momentum transfer between the hadrons\footnote{Rather than the "internal reason" by the $W$-boson mass and
the heavy $b$-quark mass in which the operator product is used at the small distance region $\overline{z_i} \ll 1/\mu_t$.},
in this case $\overline{z}_i \overline{p}_i \sim 1$ and the expansion parameter for a given operator is the twist (dimension minus spinor).
To separate the amplitude of matrix element contributed from the short- and long-distance interactions,
we replace the lines with large virtuality by the free propagators,
while retain the lines with small virtuality in the Heisenberg operator. 
In this way the matrix element can be  written in the factorizable form,
\beq
\langle \pi^-(p_2) \vert J_\mu^{\mathrm{e.m.}} \vert \pi^-(p_1) \rangle  =  && \oint dz_1 dz_2 \, \big\langle \pi^-(p_2) \bigg\vert \left\{ \overline{d}_\gamma(0) \,
\mathrm{exp} \left( ig_s \int_{z_2}^0 d\sigma_{\nu^\prime} A_{\nu^\prime}(\sigma) \right) u_\beta(z_2) \right\}_{kj} \bigg\vert 0 \big\rangle_{\mu_t}  \non
&& \, \cdot \, H_{\gamma\beta\alpha\delta}^{ijkl}(z_1,z_2) \, \cdot \, \big\langle 0 \bigg\vert \left\{ \overline{u}_\alpha(z_2) \,
\mathrm{exp} \left(i g_s \int_{z_1}^{z_2} d\sigma_\nu A_\nu(\sigma) \right) d_\delta(z_1) \right\}_{il} \bigg\vert \pi^-(p_1) \big\rangle_{\mu_t} \, ,
\label{eq:ff-fact}
\eeq
where $\gamma , \beta, \alpha, \delta$ are the spinor indices, and $i,j,k,l$ are the color indicators.
In Eq.~(\ref{eq:ff-fact}),  the hard kernel associated with the lowest Fock state is
\beq
H_{\gamma\beta\alpha\delta}^{ijkl}(z_1,z_2) = (-1) \left[ i g_s \gamma_m \right]_{\alpha\beta} T^{ij}
\left[(i e_q \gamma_\mu) S_0(0-z_1) (i g_s \gamma_n) \right]_{\gamma\delta} T^{kl}  \left[-i D_{mn}^0(z_1-z_2) \right] \,,
\label{eq:hk-2p}
\eeq
where the factor $(-1)$ comes from the anti-communicativity of the quark operator,
and the free propagators are written in the coordinate space as
\beq
S_0(z) = \frac{i}{2\pi}\frac{\zsl}{z^4} \,, \,\,\,\,\,\, D_{mn}^0(z) = \frac{1}{4\pi}\frac{g_{mn}}{z^2} \,.
\eeq
The nonlocal matrix elements in Eq.~(\ref{eq:ff-fact}) imply the amplitudes of mesons breaking-up into a pair of soft quarks,
they receive contributions from different spin structures
\beq
&&\big\langle 0 \bigg\vert \left\{ \overline{u}_\alpha(z_2) \,
\mathrm{exp} \left(i g_s \int_{z_1}^{z_2} d\sigma_\nu A_\nu(\sigma) \right) d_\delta(z_1) \right\}_{il} \bigg\vert \pi^-(p_1) \big\rangle_{\mu_t} \non
=&& \frac{\delta_{il}}{3} \left\{
\frac{1}{4} \left(\gamma_5 \gamma^\rho \right)_{\delta\alpha} \big\langle 0 \big\vert \overline{u}(z_2) \, \mathrm{exp} \left( i g_s \int_{z_1}^{z_2} \, d\sigma_\nu A_\nu(\sigma) \right) \left(\gamma_\rho \gamma_5 \right) d(z_1) \big\vert \pi^-(p_1) \big\rangle_{\mu_t} \right. \non
&& \left.  \hspace{0.5cm}+
\frac{1}{4} \left(i \gamma_5 \right)_{\delta\alpha} \big\langle 0 \big\vert \overline{u}(z_2) \, \mathrm{exp} \left( i g_s \int_{z_1}^{z_2} \, d\sigma_\nu A_\nu(\sigma) \right) \left(i \gamma_5 \right) d(z_1) \big\vert \pi^-(p_1) \big\rangle_{\mu_t} \right. \non
&& \left.  \hspace{0.5cm}+
\frac{1}{8} \left(\sigma^{\tau\tau'}\gamma_5 \right)_{\delta\alpha} \big\langle 0 \big\vert \overline{u}(z_2) \, \mathrm{exp} \left( i g_s \int_{z_1}^{z_2} \,
d\sigma_\nu A_\nu(\sigma) \right) \left(i \sigma_{\tau\tau'}\gamma_5 \right) d(z_1) \big\vert \pi^-(p_1) \big\rangle_{\mu_t} \right.\non
&& \left.  \hspace{0.5cm}+ \cdots \right\} \,.
\label{eq:Fierz}
\eeq
In the above expression, the ellipsis indicate the rest terms in the Fierz transformation,
and the truncated scale of the integral 
$\mu_t$\footnote{We will drop this indicator hereafter for the concise.} is usually known as the factorizable scale.
We quote the definition of Light-cone distribution amplitudes (LCDAs) of light pseudoscalar meson in appendix.\ref{app:DAs-definition}.

Substituting Eqs.~(\ref{eq:hk-2p},\ref{eq:Fierz} ) into Eq.~(\ref{eq:ff-fact})  and  taking into account the definition in Eq.~(\ref{eq:ff-1}),
we obtain the pion e.m. form factor at each power with the two-parton-to-two-parton scattering,
\beq
F_{\pi}^{t2}(Q^2) &=& \frac{8}{9} \alpha_s \pi f_\pi^2 Q^2 \, \int \, \frac{d^2 \mathbf{k}_{1T}}{(2\pi)^2} \,\frac{d^2 \mathbf{k'}_{1T}}{(2\pi)^2}
\int_0^1 dx \int_0^1 dy \, \varphi_\pi(x) \varphi_\pi(y) \frac{\bar{y}}{\Delta_1^2\Delta_2^2} \,,
\label{eq:ff-t2}\\
F_{\pi}^{t3,2p}(Q^2) &=& \frac{16}{9} \alpha_s \pi f_\pi^2 m_0^2 \, \int \, \frac{d^2 \mathbf{k}_{1T}}{(2\pi)^2} \,\frac{d^2 \mathbf{k'}_{1T}}{(2\pi)^2}
\int_0^1 dx \int_0^1 dy \, \frac{1}{\Delta_1^2\Delta_2^2} \non
&& \cdot  \left[ - y  \, \varphi_\pi^P(x)\varphi_\pi^P(y) - \frac{1}{6}\varphi_\pi^P(x) \varphi_\pi^\sigma(y) \left(\frac{yQ^2}{\Delta_1^2} +
\frac{(\bar{x}-\bar{y})Q^2}{\Delta_2^2} + 1 + \frac{(2-x)\bar{y}Q^2}{\Delta_2^2}\right)\right]  \,,
\label{eq:ff-t3-2p} \\
F_{\pi}^{t2 \otimes t4,2p}(Q^2) &=& \frac{16}{9} \alpha_s \pi f_\pi^2  \, \int \, \frac{d^2 \mathbf{k}_{1T}}{(2\pi)^2} \,
\frac{d^2 \mathbf{k'}_{1T}}{(2\pi)^2}  \int_0^1 dx \int_0^1 dy \, \Big\{
\frac{\bar{x}\bar{y} Q^2}{\Delta_1^2\Delta_2^4} g_{2\pi}(x) \varphi_\pi(y) \non
&&+ 2\bar{y}Q^2 \left[ \frac{1}{\Delta_1^2\Delta_2^4} + \frac{\bar{y}(2-x)Q^2}{\Delta_1^4\Delta_2^4} +
\frac{1}{\Delta_1^4\Delta_2^2} \right]  \left[ \varphi_\pi(x) g_{1\pi}(y) -\varphi_\pi(x) g ^{\dag}_{2\pi}(y) \right] \non
&&+ \left[\frac{\bar{y}^2Q^2}{\Delta_1^2\Delta_2^4} + \frac{\bar{y}^2Q^2}{\Delta_1^4\Delta_2^2} \right] \varphi_\pi(x) g_{2\pi}(y) \, \Big\} \,.
\label{eq:ff-t24-2p}
\eeq
The symbols of triangle in the above expressions represent the momentum carried by internal propagator:
$\Delta_1= \bar{y}p_2 - p_1 = (-Q/\sqrt{2}, \bar{y}Q/\sqrt{2}, \mathbf{k}')$,
$\Delta_2 = \bar{x}p_1-\bar{y}p_2 = (\bar{x}Q/\sqrt{2}, - \bar{y}Q/\sqrt{2}, \mathbf{k}-\mathbf{k}')$($\bar{y}=1-y$ and $\bar{x}=1-x$),
in which $p_1$ and $p_2$ are the momentum of initial and final pions, respectively,
$x$ and $y$ denote the momentum fraction carried by the quark in hadrons.
The twist-2 times twist-4 contribution to the form factor is studied as the first time in the PQCD approach\footnote{
Twist-4 contribution to pion form factor has been studied in the LCSRs approach,
and the result indicates a visible enhancement in the large $Q^2$ regions
which is understood by the same asymptotic behaviour $\sim 1/Q^4$ as the twist-2 contribution at $Q^2 \to \infty$\cite{BijnensMG}.}.
To obtain Eq.~(\ref{eq:ff-t24-2p}),
we have defined an auxiliary DA $g^{\dag}_2(x) \equiv \int_{0}^{x} dx' \, g_{2}(x')$ with the bound condition $g^{\dag}_2(x=0,1)=0$,
and used the following Fourier transformations,
\beq
&&\frac{1}{x^2} \Leftrightarrow -i 4\pi^2\frac{1}{p^2} \,, \,\,\,\,\,\,
\frac{x_\alpha}{x^2} \Leftrightarrow 8\pi^2\frac{p_\alpha}{(p^2)^2} \,, \,\,\,\,\,\,
\frac{x_\alpha}{(x^2)^2} \Leftrightarrow 2\pi^2\frac{p_\alpha}{p^2} \,, \,\,\,\,\,\, \non
&&\frac{x_\alpha x_\beta}{x^2} \Leftrightarrow \frac{-i 8 \pi^2}{(p^2)^2} \left(g_{\alpha\beta} - 4\frac{p_\alpha p_\beta}{p^2} \right) \,, \,\,\,\,\,\,
\frac{x_\alpha x_\beta}{(x^2)^2} \Leftrightarrow \frac{-i 2 \pi^2}{p^2} \left(g_{\alpha\beta} - 2\frac{p_\alpha p_\beta}{p^2} \right) \,.
\label{eq:fourier}
\eeq
The Sudakov exponential from $k_T$ resummation, which would be discussed in the next section,
suppresses the distribution of meson with wide transversal distance.
We can omit the transversal momenta terms on the numerator in the large momentum transferred processes,
then the second term on the right hand side of Eq.~(\ref{eq:ff-t24-2p}) vanishes, and the
contributions associated with twist-3 DAs and twist-2 times twist-4 DAs reduce to
\beq
F_{\pi}^{t3,2p}(Q^2) &\rightarrow& \frac{16}{9} \alpha_s \pi f_\pi^2 m_0^2 \, \int \, \frac{d^2 \mathbf{k}_{1T}}{(2\pi)^2} \,\frac{d^2 \mathbf{k'}_{1T}}{(2\pi)^2}
\int_0^1 dx \int_0^1 dy \, \frac{1}{\Delta_1^2\Delta_2^2} \, \non
&& \cdot\left[ - y  \, \varphi_\pi^P(x)\varphi_\pi^P(y) + \frac{1+y}{6\bar{y}}\varphi_\pi^P(x) \varphi_\pi^\sigma(y) \right]  \,,
\label{eq:ff-t3-2p-1} \\
F_{\pi}^{t2 \otimes t4,2p}(Q^2)  &\rightarrow& \frac{16}{9} \alpha_s \pi f_\pi^2  \, \int \, \frac{d^2 \mathbf{k}_{1T}}{(2\pi)^2} \,\frac{d^2 \mathbf{k'}_{1T}}{(2\pi)^2}
\int_0^1 dx \int_0^1 dy \,\frac{1}{\Delta_1^2\Delta_2^2}  \,\non
&& \cdot\left[ g_{2\pi}(x) \varphi_\pi(y) + \left( \bar{y} + \frac{\bar{y}}{\bar{x}} \right) \varphi_\pi(x) g_{2\pi}(y) \right]  \,.
\label{eq:ff-t24-2p-1}
\eeq

The gauge dependence proportional to transversal momenta in two-parton-to-two-parton scattering is
cancelled by the gauge dependence emerged in the three-parton-to-three-parton scatttering\cite{ChenPN},
then all the hard kernels in these powers hold the gauge invariance,
which in turn guarantees the $k_T$ factorization formula for the form factor up to this power correction.
We here give a short review for the gauge invariance.
Generally speaking, the Feynman diagrams of three-parton-to-three-parton scattering can be divided into four categories
by the number of the valence gluon $N_g$ attached to the internal hard gluon line.
The diagrams in category A with $N_g=0$ do not bring the gauge dependence since
they can be regarded as being from an effective lowest Fock state.
The diagrams in category B contain one valence gluon attached to hard gluon, which is the main source of gauge dependence.
Category C collect the diagrams with $N_g=2$ in which the configuration with four-gluon vertex is gauge invariant,
and the amplitudes of the other configurations with double three-gluon vertexes are also  gauge dependent.
Besides these, the diagrams with the two valence gluons scatter via a three-gluon vertex are also gauge invariant
and their amplitudes diminish by applying the Ward identity, we put them in Category D.
The gauge dependence in Categories B and C then cancel with the gauge dependence in two-parton-to-two-parton scattering
by using the equation of motion for the quark field.
It is also stated that the dominant contribution in the three-parton-to-three-parton scattering
comes from the Feynman diagram with a four-gluon vertex \cite{ChenPN}.
One of the reasons is that the nonvanishing hard kernels in other diagrams are power suppressed at least by $\mathcal{O}(1/Q)$,
which can be read directly by writing down the hard kernel for each diagrams, as did in appendix B in Ref.\cite{ChenPN} under the Feynman gauge.
Otherwise, in the PQCD approach the momentum fractions of light quarks are usually shrunk into the order $x_1,y_1 \sim \mathcal{O}(10^{-1})$
(maybe a litter larger) by the threshold resummation \cite{LiNK,ChengGBA},
a valence soft gluon attached to the internal quark propagators introduces a power suppression such as $\mathcal{O}(1/(y_1Q^2))$,
while the gluon attaches to the internal hard gluon introduces, i.e., $\mathcal{O}(1/(x_1y_1Q^2))$,
then the naive order analysis of the momentum fractions give another support.

We now consider only the gauge invariant diagram with the four-gluon vertex in three-parton-to-three-parton scattering\footnote{Two-parton-to-three-parton
and three-parton-to-two-parton scatterings are forbidden by the color transparency mechanism.},
whose contribution to the pion e.m. form factor associated with the twist-3 DAs $\varphi_{3\pi}(x_i)$ is
\beq
F_{\pi}^{t3,3p}(Q^2) = \frac{16}{3} \alpha_s \pi  f_{3\pi}^2  Q^2 \int \, \frac{\mathcal{D}^2 \mathbf{k}_{iT}}{(2\pi)^2} \,
\frac{\mathcal{D}^2 \mathbf{k'}_{iT}}{(2\pi)^2}
\int_0^1 \mathcal{D}x_i \int_0^1 \mathcal{D}y_i \, \varphi_{3\pi}(x_i) \varphi_{3\pi}(y_i) \,
\frac{1-y_1}{\varDelta_1^2 \varDelta_2^2 \varDelta_3^2} \,.
\label{eq:ff-t3-3p}
\eeq
We denote  the momenta in three-parton scattering by the oblique triangles to differentiate with the momenta in two-parton scattering:
$\varDelta_1 = p_1-p_2+k_2$, $\varDelta_2 = p_2-k_2 - (p_1-k_1)$ and $\varDelta_3 = \bar{k}_1 - \bar{k_2}$.
The momenta carried by the quark lines are $k_1=(x_1p_1^+, 0 , k_{1\perp})$ and $k_2=(0, y_1p_2^-, k'_{1\perp})$
for the initial and final mesons, respectively,
and the antiquark lines carry momenta $\bar{k}_1=(x_2p_1^+, 0 , k_{2\perp})$ and $\bar{k}_2=(0, y_2p_2^-, k'_{2\perp})$.
The integral variables  $\mathcal{D}x_i $ and   $ \mathcal{D}^2 \mathbf{k}_{iT} $  in Eq.~(\ref{eq:ff-t3-3p}) can be written in the form
\beq
\mathcal{D}x_i = dx_1dx_2dx_3 \, \delta(1-x_1-x_2-x_3),  \quad
\mathcal{D}^2\mathbf{k}_{iT} = d^2\mathbf{k}_{1T} \, d^2\mathbf{k}_{2T}.
\eeq
It is easy to see that  the contribution  $F_{\pi}^{t3,3p}(Q^2)$   is at subleading power ($\mathcal{O}(1/Q^2)$) when compared with the leading twist
contribution as given in Eq.~(\ref{eq:ff-t2}).
The contribution in the three-parton-to-three-parton scattering associated with twist-4 DAs is also firstly calculated and can be written in the
following form:
\beq
F_{\pi}^{t4,3p}(Q^2) &&= - \frac{8}{3} \alpha_s \pi f_\pi^2  \int \, \frac{\mathcal{D}^2 \mathbf{k}_{iT}}{(2\pi)^2} \,\frac{\mathcal{D}^2 \mathbf{k'}_{iT}}{(2\pi)^2}
\int_0^1 \mathcal{D}x_i \int_0^1 \mathcal{D}y_i \,
\frac{1}{\varDelta_1^2 \varDelta_2^2 \varDelta_3^2} \, \non
&&\cdot \Big\{
\varphi_\parallel^\dag(x_i) \varphi_\parallel^\dag(y_i)
\Big[ \frac{2Q^2}{\varDelta_2^2} \Big(-2(1-y_1) + \frac{y_2}{2} \Big) + \frac{5Q^2y_2}{\varDelta_3^2} + \frac{2Q^4}{(\varDelta_3^2)^2}(1-y_1)y_2x_2 \non
&&\hspace{2.2cm}+ \frac{4Q^4}{\varDelta_1^2 \varDelta_3^2} (1-y_1)[y_2+2x_2(1-y_1)] - \frac{4Q^4}{\varDelta_1^2 \varDelta_2^2}(1-y_1)^2[1+2(1-x_1)] \non
&&\hspace{2.2cm}-  \frac{2Q^4}{\varDelta_2^2 \varDelta_3^2}(1-y_1)[5(1-y_1)x_2+5(1-x_1)y_2-(1-y_1)(1-x_2)] \Big] \non
&&+ \varphi_\parallel^\dag(y_i) \varphi_\perp(x_i)  \Big[ 4 + \frac{Q^2}{\varDelta_1^2}(1-y_1) - \frac{Q^2}{\varDelta_2^2}(1-y_1)(1-x_1)
+ \frac{Q^2}{\varDelta_3^2}(1-y_1)x_2 \Big] \non
&&+ \varphi_\perp(y_i) \varphi_\parallel^\dag(x_i)   \Big[ -\frac{Q^2}{\varDelta_2^2}(1-y_1)y_1 + \frac{Q^2}{\varDelta_3^2}y_1y_2 \Big] \non
&&+ \varphi_\perp(y_i) \varphi_\perp(x_i)   \Big[ 5y_1 \Big]
+ \Big[\varphi \rightarrow \tilde{\varphi} \Big]
\Big\} \, .
\label{eq:ff-t4-3p}
\eeq
To obtain  $F_{\pi}^{t4,3p}(Q^2)$, the similar auxiliary DAs  $\varphi_\parallel^\dag(x_i)$ and $\varphi_\parallel^\dag(y_i)$  are introduced,
\beq
\varphi_\parallel^\dag(x_i) \equiv \int_0^{x_1} \, dx'_1 \, \varphi_\parallel(x'_1,x_2,x_3),  \qquad
\varphi_\parallel^\dag(y_i) \equiv \int_0^{y_2} \, dy'_2 \, \varphi_\parallel(y_1,y'_2,y_3),
\eeq
with the bound conditions $\varphi_\parallel(x_1=0/1,x_2,x_3) = 0$ and $\varphi_\parallel(y_1,y_2=0/1,y_3)=0$ , respectively.

\section{The PQCD formulae}\label{sec:PQCD}

We would like to start this section by discussing the end-point behaviours of the form factors.
The form factor at leading power $F_\pi^{t2}(Q^2)$ in Eq.~(\ref{eq:ff-t2}) does not have the end-point problem
due to the exchanging symmetry when two valence quarks form a pion in the perturbative limit.
The leading contribution with the quark-antiquark-gluon assignment $F_\pi^{t3,3p}(Q^2)$ in Eq.~(\ref{eq:ff-t3-3p})
is also end-point safe  due to  the similar reason.
The end-point problems start to emerge at the subleading power $\mathcal{O}(1/Q^2)$,
and appear in terms of the logarithm singularity  (i.e., the second term in Eq.~(\ref{eq:ff-t3-2p-1}) and the first term in Eq.~(\ref{eq:ff-t24-2p-1}) )
and the linear singularity \footnote{We thank the referee for pointing out that
the twist-2 times twist-4 contribution in Eq.~(\ref{eq:ff-t24-2p-1}) should contain only the logarithm singularity
to make sure the collinear factorization at leading twist.}  (i.e., the first term in Eq.~(\ref{eq:ff-t3-2p-1})
and the $\mathcal{O}(1/Q^4)$ correction in Eq.~(\ref{eq:ff-t4-3p}) ).
To overcome the end-point problems, we recall the transversal momentum for each external quark field
to regularize the singularity by the off-shellness $k_i^2$,
and make the resummation  for the large logarithm $\ln(Q^2/k_T^2)$  ( appeared in the high order correction to hard kernel)
to get the $k_T$ Sudakov factor,
\beq
S(x_i, y_i, b, b', \mu) = \sum_{i=1,2}\Big[ s \Big(x_i \frac{Q}{\sqrt{2}}, b\Big) +  s_q(b, \mu) \Big]
+ \sum_{i=1,2}\Big[ s \Big(y_i \frac{Q}{\sqrt{2}}, b'\Big) +  s_q(b', \mu) \Big] \,,
\label{eq:su-kt}
\eeq
where the terms $s(Q,b)$ collect the double and single logarithms in the vertex correction associated with an energetic light quark \cite{BottsKF,LiNU,CaoEQ},
and the terms $s_q(b,\mu)$  comes from the resummation of the single logarithms in the quark self-energy correction \cite{LiUN,ChengKHI},
\beq
s_q(b,\mu) 
= -\frac{1}{\beta_1} \ln \Big[ \frac{\ln (\mu/\Lambda^{(5)})}{-\ln (b\Lambda^{(5)})}\Big]
- \frac{\beta_2}{2\beta_1^3} \Big[ \frac{\ln [2\ln(\mu/\Lambda^{(5)})]+1}{\ln(\mu/\Lambda^{(5)})} - \frac{\ln [-2 \ln (b\Lambda^{(5)})]+1}{-\ln (b\Lambda^{(5)})} \Big] \,.
\label{eq:su-kt-2}
\eeq
Eq.~(\ref{eq:su-kt-2}) is obtained by considering the strong coupling at the two-loop accuracy,  $\beta_1 = (33-2 n_f)/12$ and $\beta_2 = (153-19 n_f)/24$.
We set the factorization scale at the maximal virtuality in the hard amplitude $\mu = \mathrm{Max}(1/b_1, 1/b_2, \sqrt{\bar{y}}Q)$.
The number of active quarks is chosen as
\beq
n_f(\mu) = \mathrm{Which} \, [ \, 0 < \mu < m_c, 3,  m_c \leqslant \mu < m_b, 4, m_b \leqslant \mu < m_t, 5 \,] \,,
\label{eq:nf}
\eeq
the quark pole masses are $m_c = 1.34$ GeV, $m_b=4.2$ GeV and $m_t=173$ GeV.
For the hadronic scale we take it from PDG in the $\overline{\mathrm{MS}}$ scheme \cite{TanabashiOCA}
with considering the four-loop expression of $\alpha_s$ and the three-loop matching at the quark pole masses,
\beq
&&\Lambda = \mathrm{Which} \, [ \, n_f < 3, 0.332, 3 \leqslant n_f < 4, 0.292, 4 \leqslant n_f < 5, 0.210 \,] \,.
\label{eq:lambdaqcd}
\eeq

The longitudinal momentum fractions in the initial and final state mesons also generate large logarithm
(i.e., the double logarithm $\alpha_s \ln^2 x$) in the end-point regions,
which is resumed in the convariant gauge $\partial \cdot A = 0$ to all order to produce a universal jet function \cite{StermanAJ,CataniNE,LiGI},
\beq
J(x) = - \mathrm{exp} \Big(\frac{\pi}{4} \alpha_s C_F \Big) \int_{-\infty}^{\infty} \, \frac{dt}{\pi} \, (1-x)^{\mathrm{exp}(t)} \,
\mathrm{sin} \Big( \frac{\alpha_s C_F t}{2}\Big) \, \mathrm{exp} \Big(-\frac{\alpha_s}{4\pi} C_F t^2 \Big) \,.
\label{eq:su-x}
\eeq
The jet function is factorized out from the meson wave functions and be regarded as a part of the hard kernel.
For the sake  of  simplicity, we usually adopt  the Sudakov factor  $S_t(x)$ to parameterize the jet function \cite{KeumWI,KeumPH},
\beq
S_t(x) = \frac{2^{1+2c} \, \Gamma(\frac{3}{2}+c)}{\sqrt{\pi} \, \Gamma(1+c)} \, [x(1-x)]^{c} \,.
\label{eq:su-x-para}
\eeq
This parametrization satisfies the two fundamental properties of the jet function in Eq.~(\ref{eq:su-x}) obtained by resolving the running function:
(a) it  approaches  zero  at the end-points, and  (b) it satisfies the normalization condition in the perturbative limit $\alpha_s \to 0$ ($c \to 0$).
We remark here that the threshold resummation happens only for the high twist contributions,
and the jet function modifies the shapes of  the high twist LCDAs, especially in the end-point region,
to be proportional to $x(1-x)$ ( as parameterized in Eq.~(\ref{eq:su-x-para}) ), which then eliminates effectively the end-point singularity.

Considering the next-to-leading-order (NLO) QCD correction,
the mixed logarithm $\ln (\zeta^2 /k^2_T) \ln x$ appears in the transversal-momentum-dependent (TMD) pion wave function
\footnote{ Recently, a nondipolar gauge link for the TMD pion wave function is suggested \cite{LiXDA,WangQQR}
to eliminate the pinched singularity in the self-energy correction of non-light-like Wilson line.
This new definition is much simpler than the long-standing dipolar Wilson lines with a complicated soft subtraction definition \cite{Collins-TMD}.
In this work we would not deal with the pinched singularity problem
because the NLO pion wave function with the nondipole definition is still missing at subleading twist. },
and the variable $\zeta^2 \equiv 4 (p \cdot n)^2/n^2$  ( $p$ is the meson momenutm, $n$ is a vector deviated lightly from the light-cone $n^2 \neq 0$ )
brings the scheme dependence on a typical choice of Wilson line.
The joint resummation with off-shell Wilson line has been proposed to resolve this problem,
and the joint-resummed TMD pion wave function highlights the moderate $x$ and small $b$ regions for the momentum distribution \cite{LiXNA},
as an supplement to the conventional $k_T$ and threshold resummations.
Considering the complicated expression of  the joint-resummed wave function brings an minor impact on the pion form factor,
in this work we would still adopt the conventional pion wave function to estimate the different power contributions with setting $\zeta^2 = Q^2$.

The formulas in Eqs.(\ref{eq:su-kt},\ref{eq:su-x}) are derived specially for the two-parton-to-two-parton scattering,
and they are not available any more for the three-parton-to-three-parton scattering
since the Sudakov factor associated with a valence gluon must differ from that associated with a valence quark.
To evade the Sudakov factor for the valence gluon which is still missing in the factorization theorem,
we consider only the effective Sudakov factor associated with the most energetic quarks in the quark-antiquark-gluon Fock state,
and neglect the Sudakov factors associated with the gluon and the soft quarks \cite{ChenPN}.
The approximation is taken as\footnote{In fact,
$\mathbf{b}_2 = \mathbf{b'}_2$ due to the Gaussion integral in Eq.~(\ref{eq:propa-trans-3p}).},
\beq
S^{3}(x_i, y_i, b_i, \mu) &=& s \Big((1-x_1) \frac{Q}{\sqrt{2}}, b_1\Big) + s \Big(x_2 \frac{Q}{\sqrt{2}}, b_2\Big)
+ s\Big((1-y_1) \frac{Q}{\sqrt{2}}, b'_1\Big) + s \Big(y_2 \frac{Q}{\sqrt{2}}, b'_2\Big) \,,
\label{eq:su-kt-3p}
\eeq
and the factorization scale is modified to
\beq
\mu = \mathrm{Max}[1/b_1, 1/b_2, 1/{b'_1},  \sqrt{(1-y_1)}Q] \,.
\label{eq:scale-1}
\eeq
For the transversal component of the momentum integral, it is more convenient to do in the coordinate space,
and the Fourier transformation with two propagators reads
\beq
&~&\int d\mathbf{b}^2_1 \, d\mathbf{b'}^2_{1} \, \mathrm{exp} \Big( -i \mathbf{k}_1 \cdot \mathbf{b}_1 - i \mathbf{k'}_1 \cdot \mathbf{b'}_1 \Big)
\, \int \frac{ \, d^2\mathbf{k}_{1T}}{(2\pi)^2}\, \frac{ \, d^2\mathbf{k'}_{1T}}{(2\pi)^2}
\frac{1}{\alpha + \mathbf{k'}_1^{2}} \, \frac{1}{\beta + (\mathbf{k'}_1-\mathbf{k}_1)^2} \non
&=& \int_0^\infty \, b_1 db_1 b'_1 db'_1 \, K_0(\sqrt{\beta} b'_1) \Big[ \Theta(b_1-b'_1) I_0(\sqrt{\alpha} b'_1) \, K_0(\sqrt{\alpha} b_1) - [b_1 \leftrightarrow b'_1] \Big] \,.
\label{eq:propa-trans-2p}
\eeq
$I_0$ and $K_0$ are the  modified Bessel functions of the first and second kind, respectively, $K_0$ is also called as Basset function.
For the contribution with three internal propagators, the transversal integral is revised to
\beq
&~&\int d\mathbf{b}^2_1 \, d\mathbf{b'}^2_{1} \,d\mathbf{b}^2_2 \, d\mathbf{b'}^2_{2}
\, \mathrm{exp} \Big( -i \mathbf{k}_1 \cdot \mathbf{b}_1 - i \mathbf{k'}_1 \cdot \mathbf{b'}_1 -i \mathbf{k}_2 \cdot \mathbf{b}_2 - i \mathbf{k'}_2 \cdot \mathbf{b'}_2 \Big) \non
&~& \cdot \int \frac{ \, d^2\mathbf{k}_{1T}}{(2\pi)^2}\, \frac{ \, d^2\mathbf{k'}_{1T}}{(2\pi)^2}\, \frac{ \, d^2\mathbf{k}_{2T}}{(2\pi)^2}\, \frac{ \, d^2\mathbf{k'}_{2T}}{(2\pi)^2}
\frac{1}{\alpha + \mathbf{k'}_1^2} \, \frac{1}{\beta + (\mathbf{k'}_1-\mathbf{k}_1)^2} \, \frac{1}{\gamma + (\mathbf{k'}_2-\mathbf{k}_2)^2} \non
&=& \int_0^\infty \,b_1 db_1 b'_1 db'_1 \, K_0(\sqrt{\beta} b'_1) \Big[ \Theta(b_1-b'_1) I_0(\sqrt{\alpha} b'_1) \, K_0(\sqrt{\alpha} b_1) - [b_1 \leftrightarrow b'_1] \Big] \,
\int_0^\infty \, b_2^2 db_2 \, K_0(\sqrt{\gamma} b_2)  \,.
\label{eq:propa-trans-3p}
\eeq

In the past twenty years, the PQCD factorization approach has made many progresses in the calculation for the NLO QCD corrections
\footnote{Besides the NLO QCD corrections, the power correction with high twist distribution amplitudes is also studied \cite{ShenVDC}.}.
Here we give a brief summary about the major progresses for light meson form factors.
The NLO calculation for pion e.m form factor associated with two-parton twist-2 and twist-3 DAs are carried out in
Ref. \cite{LiNK} and Ref. \cite{ChengGBA}, respectively,
following which, the NLO correction to timelike pion form factor is obtained by the analytical continuum technology \cite{HuCP,ChengQRA}.
Another important correction is for the scalar pion form factor appeared in the factorizable annihilation diagrams \cite{ChengRKA},
which provides the dominate strong phase in PQCD approach to deal with two-body nonleptonic charmless $B$ decays.
Recently the NLO calculation has been done  for the $\rho\pi$ transition process to determine the strong coupling $g_{\rho\pi\gamma}$ \cite{HuaKHO},
and for the $\rho$ form factors \cite{ZhangBHJ}.
All the calculations turn out that the convergency of perturbative expansion is good in the considered energy regions,
which examines the prediction power of PQCD at the NLO level.
We would include the QCD corrections in the following numerical analysis for the two-parton-to-two-parton scattering,
and here we quote the NLO correction functions \cite{LiNK,ChengGBA},
\beq
F_{t2}^{(1)} (x_i, t, Q^2) &=& \frac{\alpha_s C_F}{4\pi} \Big[ -\frac{3}{4}\ln\frac{t^2}{Q^2} - \ln^2x_1- \ln^2x_2 + \frac{45}{8}\ln x_1\ln x_2 + \frac{5}{4} \ln x_1 + \frac{77}{16} \ln x_2 \non
&& \hspace{1cm}+ \frac{\ln 2}{2} + \frac{5}{48}\pi^2 + \frac{53}{4} \Big] \,,
\label{eq:ff-nlo-t2} \\
F_{t3}^{(1)} (x_i, t, Q^2) &=& \frac{\alpha_s C_F}{4\pi} \Big[ - \frac{9}{4}\ln\frac{t^2}{Q^2} - \frac{53}{16} \ln (x_1x_2) - \frac{1}{8} \ln^2x_2  - \frac{23}{16}\ln x_1- \frac{2}{9} \ln x_2 \non
&& \hspace{1cm}- \frac{137}{96} \pi^2 + \frac{\ln 2}{4} + \frac{337}{64} \Big] \,.
\label{eq:ff-nlo-t3-1}
\eeq

\section{Numerical results}\label{sec:numerics}

The contributions to the pion form factor from  the two-parton-to-two-parton scattering and the three-parton-to-three-parton scattering are 
rewritten compactly as the following forms,
\beq
F^{2p}_\pi(Q^2) = &&\frac{8}{9} \alpha_s \pi f_\pi^2 Q^2 \int_0^1 dx \int_0^1 dy \, \int_0^{1/\Lambda} \, b_1 db_1 b'_1 db'_1 \, e^{-S(x_i,y_i,b,b',\mu)}  \non
&& \cdot \Big\{ \bar{y} \varphi_\pi(x) \varphi_\pi(y) \, \Big[1+ F_{t2}^{(1)} (x, y, t, Q^2)\Big] \, \mathcal{H}\non
&& + \frac{2m_0^2}{Q^2} \Big[- y \varphi_\pi^P(x) \varphi_\pi^P(y) \left[1+F_{t3}^{(1)} (x, y, t, Q^2)\right]  \, \mathcal{H} \non
&& \hspace{1.2cm} + \frac{1}{6} \varphi_\pi^P(x) \varphi_\pi^\sigma(y) \left[-y Q^2\, \mathcal{H}_1
-(\bar{x}-\bar{y}-2\bar{y}+x\bar{y})Q^2 \, \mathcal{H}_2  - 1\right]  \Big] \, S_t(\bar{y})\non
&& + \frac{2}{Q^2} \Big[  \, g_{2\pi}(x) \varphi_\pi(y) \, \bar{x}\bar{y}  Q^2 \, \mathcal{H}_2
+ \varphi_\pi(x) g_{2\pi}(y)  \, \bar{y}^2 Q^2 \,  \left[\mathcal{H}_1 + \mathcal{H}_2 \right] \non
&& \hspace{0.6cm} + \big[ \varphi_\pi(x) g_{1\pi}(y)  - \varphi_\pi(x) g_{2\pi}^{\dag}(y) \big]
\left[ 2 \bar{y}Q^2( \mathcal{H}_1 + \mathcal{H}_2 + \bar{y}(2-x)Q^2 \, \mathcal{H}_3) \right]\Big]  \, S_t(\bar{y})\Big\} \,,
\label{eq:ff-2p} \\
F^{3p}_\pi(Q^2) = &&\frac{16}{3} \alpha_s \pi f_{\pi}^2 Q^2 \int_0^1 \mathcal{D}x_i \int_0^1 \mathcal{D}y_i \,
\int_0^{1/\Lambda} \, b_1 db_1 b'_1 db'_1 b_2^2 db_2 \, e^{-S^3(x_i,y_i,b_i,\mu)}  \non
&& \cdot \Big\{ \frac{f_{3\pi}^2}{f_{\pi}^2} (1-y_1) \varphi_{3\pi}(x_i) \varphi_{3\pi}(y_i) \, \mathcal{H}' \non
&& + \frac{1}{2Q^2} \Big[
\varphi_\parallel^\dag(x_i) \varphi_\parallel^\dag(y_i)
\big[ \left(-4 (1-y_1)+ (1-y_1)y_2 \right) Q^2 \, \mathcal{H}'_2 + 5(1-y_1)y_2Q^2 \,\mathcal{H}'_3 \big] \non
&&\hspace{0.7cm} + \varphi_\parallel^\dag(y_i) \varphi_\perp(x_i)  \big[ 4\,\mathcal{H}' +(1-y_1)Q^2 \, \mathcal{H}'_1 - (1-x_1)(1-y_1)Q^2 \, \mathcal{H}'_2 + (1-x_1)(1-y_1)x_2 Q^2 \, \mathcal{H}'_3 \big] \non
&&\hspace{0.7cm} + \varphi_\perp(y_i) \varphi_\parallel^\dag(x_i)  \big[ -y_1(1-y_1)Q^2\, \mathcal{H}'_2 + y_1 (1-y_1) y_2Q^2 \, \mathcal{H}'_3 \big] \non
&&\hspace{0.7cm} + \varphi_\perp(y_i) \varphi_\perp(x_i)   5y_1 \, \mathcal{H}'
+ [\varphi \rightarrow \tilde{\varphi} ]  \Big] \Big\} \,.
\label{eq:ff-3p}
\eeq
The hard functions appeared in Eq.~(\ref{eq:ff-2p}) and Eq.~(\ref{eq:ff-3p}) can be  written in terms of the Bessel functions,
\beq
&&\mathcal{H}(\alpha, \beta, b_1,b'_1) = K_0(\sqrt{\beta} b'_1) \Big[ \Theta(b_1-b'_1) I_0(\sqrt{\alpha} b'_1) \, K_0(\sqrt{\alpha} b_1) - [b_1 \leftrightarrow b'_1] \Big]  \,,
 \label{eq:hf-2p-1}\\
&&\mathcal{H}_1(\alpha, \beta, b_1,b'_1) = K_0(\sqrt{\beta} b'_1) \Big[ \frac{b_1}{2\sqrt{\alpha}}  \Theta(b_1-b'_1) I_0(\sqrt{\alpha} b'_1) \, K_1(\sqrt{\alpha} b_1) - [b_1 \leftrightarrow b'_1] \Big] \,,
 \label{eq:hf-2p-2}\\
&&\mathcal{H}_2(\alpha, \beta, b_1,b'_1) = \frac{b'_1}{2\sqrt{\beta}} K_1(\sqrt{\beta} b'_1) \Big[ \Theta(b_1-b'_1) I_0(\sqrt{\alpha} b'_1) \, K_0(\sqrt{\alpha} b_1) - [b_1 \leftrightarrow b'_1] \Big]  \,,
 \label{eq:hf-2p-3}\\
&&\mathcal{H}_3(\alpha, \beta, b_1,b'_1) = \frac{b'_1}{2\sqrt{\beta}} K_1(\sqrt{\beta} b'_1) \Big[ \frac{b_1}{2\sqrt{\alpha}} \Theta(b_1-b'_1) I_0(\sqrt{\alpha} b'_1) \, K_1(\sqrt{\alpha} b_1) - [b_1 \leftrightarrow b'_1] \Big]  \,,
 \label{eq:hf-2p-4}
\eeq
\beq
&&\mathcal{H}' (\alpha', \beta', \gamma, b_1,b'_1,b_2) = K_0(\sqrt{\gamma} b_2) K_0(\sqrt{\beta'} b'_1) \Big[ \Theta(b_1-b'_1) I_0(\sqrt{\alpha'} b'_1) \, K_0(\sqrt{\alpha'} b_1) - [b_1 \leftrightarrow b'_1] \Big]  \,,
 \label{eq:hf-3p-1}\\
&&\mathcal{H}'_1(\alpha', \beta', \gamma, b_1,b'_1,b_2) = K_0(\sqrt{\gamma} b_2) K_0(\sqrt{\beta'} b'_1) \Big[ \frac{b_1}{2\sqrt{\alpha'}} \Theta(b_1-b'_1) I_0(\sqrt{\alpha'} b'_1) \, K_1(\sqrt{\alpha'} b_1) - [b_1 \leftrightarrow b'_1] \Big]  \,,
\label{eq:hf-3p-2} \\
&&\mathcal{H}'_2(\alpha', \beta', \gamma, b_1,b'_1,b_2) = \frac{b'_1}{2\sqrt{\beta'}} K_0(\sqrt{\gamma} b_2) K_1(\sqrt{\beta'} b'_1) \Big[ \Theta(b_1-b'_1) I_0(\sqrt{\alpha'} b'_1) \, K_0(\sqrt{\alpha'} b_1) - [b_1 \leftrightarrow b'_1] \Big]  \,,
 \label{eq:hf-3p-3}\\
&&\mathcal{H}'_3(\alpha', \beta', \gamma, b_1,b'_1,b_2) = \frac{b_2}{2\sqrt{\gamma}} K_1(\sqrt{\gamma} b_2) K_0(\sqrt{\beta'} b'_1) \Big[ \Theta(b_1-b'_1) I_0(\sqrt{\alpha'} b'_1) \, K_0(\sqrt{\alpha'} b_1) 
- [b_1 \leftrightarrow b'_1] \Big]  \,.
\label{eq:hf-3p-4}
\eeq
To obtain the above expressions, we have defined the following denotation for the internal virtuality,
\beq
&&\alpha \equiv \bar{y} Q^2 \,, \quad  \beta \equiv \bar{x}\bar{y}Q^2 \,,  \non
&&\alpha' \equiv (1-y_1) Q^2 \,, \quad  \beta' \equiv (1-x_1)(1-y_1)Q^2 \,, \quad \gamma \equiv x_2y_2Q^2 \,.
\label{eq:vaituality}
\eeq
For the form factor of kaon, we simply make  the replacements $f_\pi \rightarrow f_K, \, m_0^\pi \rightarrow m_0^K$
and also for the nonperturbative parameters in meson DAs.
The power expansion is shown explicitly in Eqs.~(\ref{eq:ff-2p},\ref{eq:ff-3p}),
which reads $\mathcal{O}(1) : \mathcal{O}(\frac{m_0^2}{Q^2}) : \mathcal{O}(\frac{\delta_P^2}{Q^2}) :
\mathcal{O}(\frac{f_{3\mathcal{P}}^2}{f_{\mathcal{P}}^2 Q^2}) : \mathcal{O}(\frac{\delta_{\mathcal{P}}^4}{Q^4})$
corresponding to the contributions associated with leading twist,
two-parton twist-3, twist-2 times twist-4, three-parton twist-3 and twist-4 DAs, respectively.

\begin{table}[b]
\caption{Hadronic parameters for $\pi$ and $K$ meson DAs in our evaluation.}
\begin{center}
\begin{tabular}{|c|c||c|c||c|}\hline\hline
$\pi$ & $\mu = 1 \, \mathrm{GeV}$ & $K$ & $\mu = 1 \, \mathrm{GeV}$ & $\mathrm{Remarks/Refs}$ \\   \hline
$f_\pi$ & $0.13$ & $f_K$ & $0.16$ & $\mathrm{in \, unit \, of \, GeV}$, \cite{TanabashiOCA}  \non
$m_0^\pi$ & $1.9$ & $m_0^K$ & $1.9$ & $\mathrm{in \, unit \, of \, GeV}$, \cite{LeutwylerQG,KhodjamirianYS}  \non
\hline
$a_1^\pi$ & $0$ & $a_1^K$ & $0.064 \pm 0.0041$ & \cite{BaliDQC}  \non
$a_2^\pi$ & $0.13 \pm 0.028$ & $a_2^K$ & $0.12 \pm 0.025$ & $a_{n \geqslant 2} = 0$, \cite{BaliDQC}  \non
\hline
$f_{3\pi}$ & $\,\,\,0.0045\pm 00015$ \,\,\,& $f_{3K}$ & \,\,\,$0.0045 \pm 0.0015$ \,\,\,& \,\,\,$\mathrm{in \, unit \, of \, GeV}^2$, \cite{BraunIV,BallWN} \non
$\omega_{3\pi}$ & $-1.5 \pm 0.7$ & $\omega_{3K}$ & $-1.2 \pm 0.7$ &  \cite{BallWN} \non
$\lambda_{3\pi}$ & $0$ & $\lambda_{3K}$ & $1.6 \pm 0.4$ & \cite{BallWN} \non
\hline
$\delta_\pi^2$& $0.18 \pm 0.06$ & $\delta_K^2$ &$0.20 \pm 0.06$ & \,\,\,$\mathrm{in \, unit \, of \, GeV}^2$, \cite{BallWN,NovikovJT,BakulevUC} \non
$\omega_{4\pi}$& $0.20 \pm 0.10$ & $\omega_{4K}$ & $0.20\pm0.10$ & \cite{BallWN,KhodjamirianYS} \non
$\kappa_{4\pi}$ & $0$ &$\kappa_{4K}$ & $-0.12\pm0.01$ & \cite{BallWN,KhodjamirianYS}\\ \hline
\end{tabular}
\end{center}
\label{tab-I}
\end{table}

We take the PDG value $\overline{m}_s(2 \, \mathrm{GeV}) = 96^{+8}_{-4} \, \mathrm{MeV}$
corresponding to $\overline{m}_s(1 \, \mathrm{GeV}) =125^{+10}_{-5} \, \mathrm{MeV}$.
The well-known chiral perturbative theory (ChPT) relations \cite{LeutwylerQG}
\beq
\mathcal{R} \equiv \frac{2m_s}{m_u+m_d} = 24.4 \pm 1.5 \,, \,\,\,\,\,\, \mathcal{Q}^2 \equiv \frac{m_s^2-(m_u+m_d)^2/4}{m_d^2-m_u^2} = (22.7 \pm 0.8)^2
\label{eq:ChPT-relation}
\eeq
is used to determine the chiral masses of light mesons
\beq
m_0^\pi = \frac{m_\pi^2 \mathcal{R}}{2m_s} \,, \,\,\,\,\,\,
m_0^K = \frac{m_K^2}{m_s \left[1 + \frac{1}{\mathcal{R}} \left( 1- \frac{\mathcal{R}^2-1}{4\mathcal{Q}^2}\right)\right]} \,,
\eeq
without involving the light quark masses $m_u$ and $m_d$ because we neglect them elsewhere besides in $m_0^\pi$ and $m_0^K$.
The parameters for meson DAs chosen for the numerical evaluation are listed  in Table.\ref{tab-I},
in which  the Gegenbauer moments $a_2^\pi, a_1^K, a_2^K$ are evaluated from LQCD
with the new developed momentum smearing technique \cite{BaliUDE},
all others are calculated from QCD sum rules\footnote{Recently,
the feasibility of calculating the pion DAs from suitably chosen Euclidean correlation functions at large momentum is investigated,
this method allow us to study higher-twist DAs from LQCD\cite{BaliGFR, BaliSPJ},
and the result for the parameter $\delta_\pi^2$ consists with it estimated from QCD sum rules,
even though the systematic errors is still not yet under control. }.

Our prediction of pion and kaon form factors is illustrated in Figure.~\ref{fig:1}, where the contributions from different powers are shown separately.
The contributions at leading (Red  dashed-curves) and subleading twists (Blue dotted-curves) with two-parton-to-two-parton scattering
have been included the the NLO QCD corrections \cite{LiNK,ChengGBA}.
The chiral enhancement at twist-3 is shown evidently, and this effect for kaon form factor is stronger than that for the pion form factor.
We define a ratio between the subleading and the leading twist contributions as $\mathrm{R_P(Q^2)  \equiv F_P^{T2}(Q^2)/F_P^{T3-2P}(Q^2)}$
with the notation $\mathrm{P} = \pi$ and K,
and take the deviation of their relative magnitude from the unit $\mathrm{A \equiv 1 - R_\pi(Q^2)/R_K(q^2)}$ to estimate the $\mathrm{SU(3)}$ asymmetry.
The result shows that this asymmetry does not exceed $30 \%$ in the considered energy region and vanishes in the perturbative limit.
Figure.~\ref{fig:1} also indicates explicitly the power behavior as we claimed below Eq.~(\ref{eq:vaituality}):
the contributions from three-parton Fock states is at least one order lower than the leading contribution from lowest Fock state
in the larger energy regions $Q^2 \geqslant 10 \mathrm{GeV}^2$,
while the twist-2 times twist-4 contribution in the two-parton-to-two-parton scattering
is a litter bit larger than the contribution from three-parton-to-three-parton scattering, but they are still in the same order.

\begin{figure}[!tb]
\begin{center}
\vspace{1cm}
\includegraphics[width=0.4\textwidth]{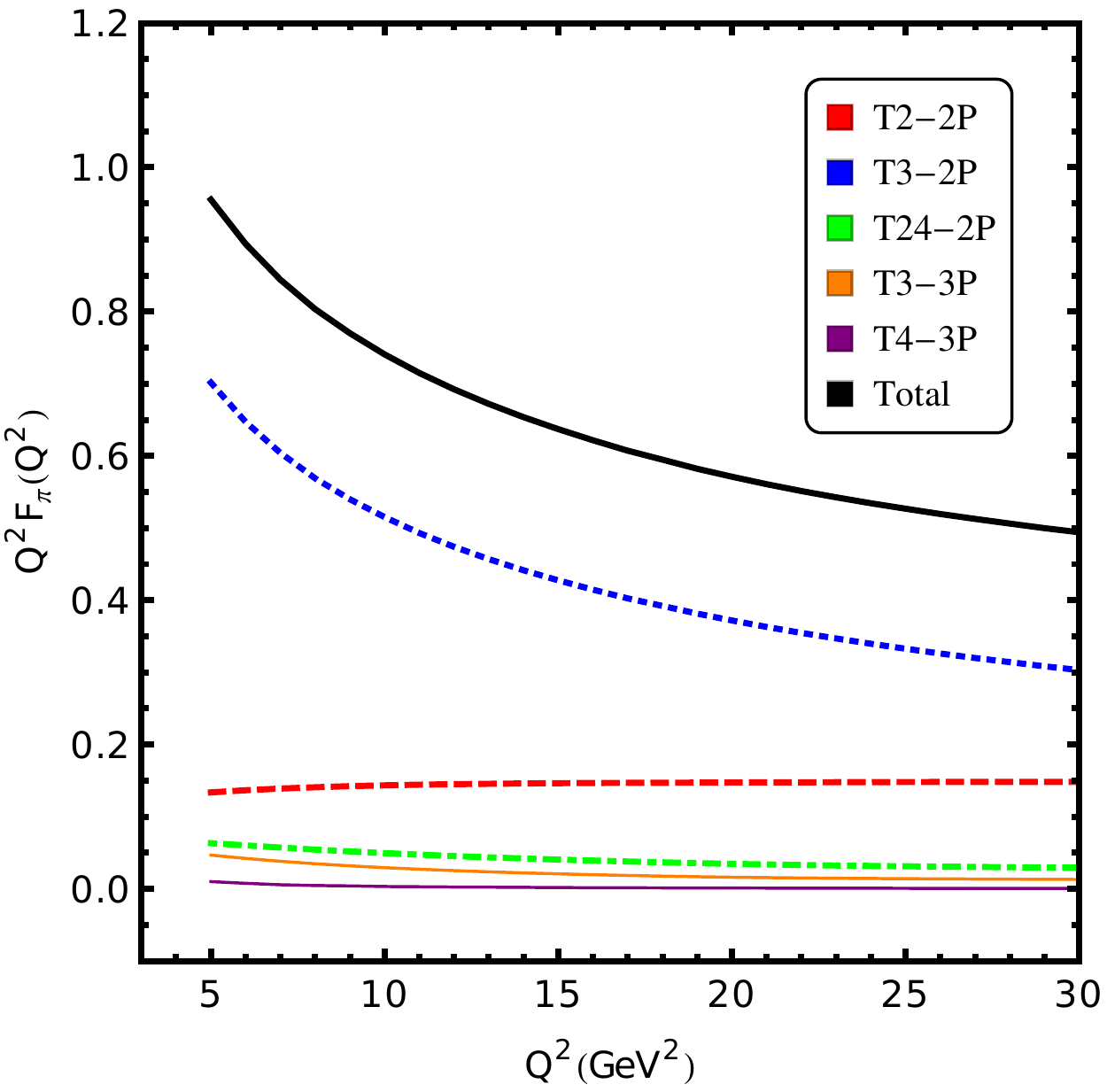}
\hspace{4mm}
\includegraphics[width=0.4\textwidth]{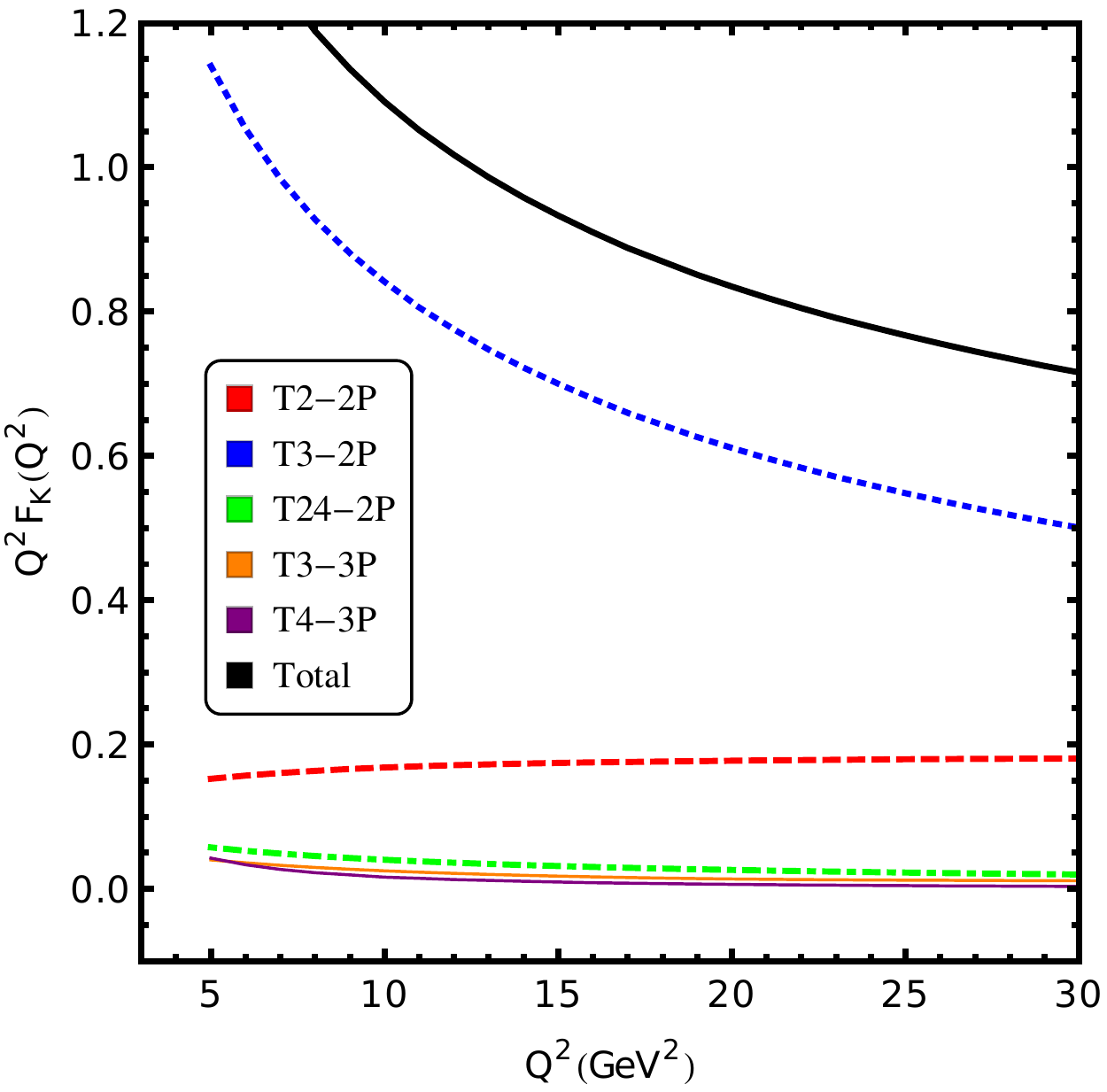}
\end{center}
\vspace{-1cm}
\caption{{\footnotesize Pion (left) and Kaon (right) form factors calculated in the PQCD approach.}}
\label{fig:1}
\end{figure}

As listed in Table \ref{tab-II}, we compare  our PQCD predictions  with the LCSRs  results \cite{ff-pion-LCSRs,BijnensMG}
at the energy point $Q^2 = 10 \, \mathrm{GeV}^2$, where the theoretical error in our calculation mainly comes from the input of the DAs,
the two sources of uncertainty in LCSRs approach are the DAs inputs and the parameters of the approach itself.
The choice of the scale for the nonperturative parameters affects weakly in the larger energy regions so we do not consider it here.
We find that the prediction of the pion and kaon form factors is comparable in the chosen energy point within the uncertainty,
and the difference between  the numerical results obtained in these two approaches becomes smaller when $Q^2$ is increasing.

\begin{table}[tb]
\caption{The PQCD and LCSRs predictions for the values of $Q^2F_{\pi,K}(Q^2)$ at the point $Q^2 = 10 \, \mathrm{GeV}^2$.}
\begin{center}
\begin{tabular}{|c||c|c||c|c|} \hline
$Q^2(\mathrm{GeV^2})$ & \,$Q^2F^{\mathrm{PQCD}}_\pi(Q^2)$ \,& \,$Q^2F_\pi^{\mathrm{LCSRs}}(Q^2)$ \,
& \, $Q^2F_K^{\mathrm{PQCD}}(Q^2)$ \, & \,$Q^2F^{\mathrm{LCSRs}}_K(Q^2)$ \, \non
\hline
 $10$ & $0.75(10)  $ & $0.51(15)$ & $1.08(15)$ & $0.76(22)$ \non
 \hline
\end{tabular}
\end{center}
\label{tab-II}
\end{table}

\section{Conclusion}\label{sec:conclusion}

We study the pion and kaon electromagnetic form factors with the inclusion of  the high power contributions up to twist-4 of the meson DAs,
the PQCD calculation confirms the convergence behaviour of  the twist expansion,
which shows that the contribution from the three-parton Fock state is at least one order of magnitude smaller than that  from the lowest Fock state.
The chiral enhancement of the subleading power contribution depends strongly on the corresponding DAs,
and this effect is quite obvious in our choice of the conformal expansion of twist-3 DAs.
The direct comparison between  the contributions to the pion and kaon form factors from  the two-parton-to-two-parton scattering indicates
that the $\mathrm{SU(3)}$ asymmetry is no more than $30 \%$ in the considered energy region.
Because the current lattice QCD evaluation and experiment measurement of the meson form factors are still in the small $Q^2$ region,
our calculation can not interplay directly with them now,
we look forward to see more data in the intermediate energy regions at Jefferson Lab with the $12 \, \mathrm{GeV}$ upgrade program,
with which the precise PQCD predictions presented in this paper can be forwarded to extract the nonperturbative parameters of meson DAs,
i.e., the moments in Gegenbauer expansion.
We compare our results with the predictions from the LCSRs approach at the fixed energy point,
and find  the parallel prediction power of these two approaches.
The further improvement in this project is to combine the precise measurement of the time-like pion and kaon form factors
in the resonance energy regions with the PQCD calculation at the large energy regions,
in order to determine the meson distribution amplitudes.

\section{ACKNOWLEDGEMENTS}

We are grateful to Hsiang-nan Li, Yu-ming Wang, Zhen-jun Xiao and Yi-bo Yang for helpful discussions,
and especially to Hsiang-nan Li and Zhen-jun Xiao for the careful reading of the manuscript.
This work is supported by the National Science Foundation of China under No. 11805060
and "the Fundamental Research Funds for the Central Universities" under No. 531118010176.

\begin{appendix}

\section{Definition of the distribution amplitudes}\label{app:DAs-definition}

Light-cone distribution amplitudes (LCDAs) for pseudoscalar meson with quark-antiquark assignment
is defined by the nonlocal matrix element sandwiched between the meson state and vacuum \cite{BallWN,BallJE},
\beq
&&\big\langle 0 \big\vert \overline{u}(z_2) (\gamma_\rho \gamma_5) q(z_1) \big\vert \mathcal{P}^-(p) \big\rangle \non
=&& f_\mathcal{P} \int_0^1 dx \, e^{-i xpz_1 - i\overline{x}pz_2}
\Big\{ i p_\rho \left[ \varphi_\mathcal{P}(x) + (z_1-z_2)^2 g_{1\mathcal{P}}(x) \right] + \left[ (z_1-z_2)_\rho - \frac{p_\rho (z_1-z_2)^2}{p(z_1-z_2)} \right] g_{2\mathcal{P}}(x) \Big\}  \,,
\label{eq:DAs-A}\\
&&\big\langle 0 \big\vert \overline{u}(z_2) (\sigma_{\tau\tau'} \gamma_5) q(z_1) \big\vert \mathcal{P}^-(p) \big\rangle \non
=&& f_\mathcal{P} m_0^\mathcal{P} \int_0^1 dx \,  e^{-i xpz_1 - i\overline{x}pz_2} \left(1 - \frac{m_\pi^2}{(m^0_\mathcal{P})^2}\right)
\left[ p^\tau (z_1-z_2)_{\tau'} - p^{\tau'}(z_1-z_2)_\tau \right] \, \varphi_\mathcal{P}^\sigma(x) \,,
\label{eq:DAs-T}\\
&&\big\langle 0 \big\vert \overline{u}(z_2) (i \gamma_5) q(z_1) \big\vert \mathcal{P}^-(p) \big\rangle
= f_\mathcal{P} m_0^\mathcal{P} \int_0^1 dx \,  e^{-i xpz_1 - i\overline{x}pz_2} \varphi_\mathcal{P}^P(x) \,,
\label{eq:DAs-P}
\eeq
where $f_\mathcal{P}$ is the decay constant, $m_0^\mathcal{P}$ is the chiral mass of  the pseudoscalar meson,
$\varphi_\mathcal{P}$, $\varphi^{P, \sigma}_\mathcal{P}$ and $g_{1\mathcal{P},2\mathcal{P}}$ corresponds to the DAs at twist-2, twist-3 and twist-4, respectively.

For the quark-antiquark-gluon assignment, the DAs are defined with the matrix element with the gluon field strength tensor operator
$G_{\kappa\kappa'}=g_s G_{\kappa\kappa'}^a \lambda^a/2$,
\beq
&&p^+ \big\langle 0 \big\vert \overline{u}(z_2) (\sigma_{\tau\tau'} \gamma_5) G_{\kappa\kappa'}(z_0) q(z_1) \big\vert \mathcal{P}^-(p) \big\rangle \non
=&& i f_{3\mathcal{P}} \int \mathcal{D}x_i \, e^{-i x_1pz_1 - ix_2pz_2 -ix_3z_0}
\left[ \left(p_{\kappa}p_{\tau} g_{\kappa'\tau'} - p_{\kappa'} p_{\tau} g_{\kappa\tau'} \right)
- \left( p_{\kappa}p_{\tau'} g_{\kappa'\tau} - p_{\kappa'} p_{\tau'} g_{\kappa\tau}\right)\right] \varphi_{3\mathcal{P}}(x_i) \,,
\label{eq:DAs-3p-T}\\
&&p^+ \big\langle 0 \big\vert \overline{u}(z_2) (\gamma_\rho \gamma_5) G_{\kappa\kappa'}(z_0) q(z_1) \big\vert \mathcal{P}^-(p) \big\rangle \non
=&& f_{\mathcal{P}} \int \mathcal{D}x_i \, e^{-i x_1pz_1 - ix_2pz_2 -ix_3z_0}
\left[ p_\rho \frac{p_{\kappa}(z_1-z_2)_{\kappa'} - p_{\kappa'}(z_1-z_2)_{\kappa}}{p(z_1-z_2)} \varphi_\parallel(x_i)
+ (g^{\perp}_{\rho\kappa}p_{\kappa'} - g^{\perp}_{\rho\kappa'} p_\kappa) \varphi_\perp(x_i) \right] \,,
\label{eq:DAs-3p-A} \\
&&p^+ \big\langle 0 \big\vert \overline{u}(z_2) (\gamma_\rho ) \tilde{G}_{\kappa\kappa'}(z_0) q(z_1) \big\vert \mathcal{P}^-(p) \big\rangle \non
=&& f_{\mathcal{P}} \int \mathcal{D}x_i \, e^{-i x_1pz_1 - ix_2pz_2 -ix_3z_0}
\left[ p_\rho \frac{p_{\kappa}(z_1-z_2)_{\kappa'} - p_{\kappa'}(z_1-z_2)_{\kappa}}{p(z_1-z_2)} \tilde{\varphi}_\parallel(x_i)
+ (g^{\perp}_{\rho\kappa}p_{\kappa'} - g^{\perp}_{\rho\kappa'} p_\kappa) \tilde{\varphi}_\perp(x_i) \right] \,,
\label{eq:DAs-3p-A}
\eeq
where $\tilde{G}_{\kappa\kappa'}=1/2 \, \epsilon_{\kappa\kappa'\eta\eta'} G^{\eta\eta'}$,
the location of gluon file strength is at $z_0=vz_1+\overline{v}z_2$ with the free variable $v \in [0,1]$,
$\varphi_{3\mathcal{P}}$ is the twist-3 DA, and $\varphi_{\parallel,\perp}, \tilde{\varphi}_{\parallel,\perp}$ are twist-4 DAs.
When $q=d, s$, the meson $\mathcal{P}= \pi, K$, respectively.

\section{Expressions of the distribution amplitudes}\label{app:DAs-expression}

LCDAs can be obtained by using the conformal partial expansion,
and the most familiar expression is the leading twist DAs written in terms of the Gegenbauer polynomials,
\beq
\varphi_{\mathcal{P}}(x, \mu) = 6x\bar{x} \sum_{n=0} \, a_n(\mu)  \, C_n^{3/2}(2x-1) \,.
\label{eq:DA-t2}
\eeq

Two-particle twist-3 DAs are related to the three-particle DA $\varphi_{3\mathcal{P}}(x_i)$ and
also to the leading twist DA $\varphi_\mathcal{P}$ by the QCD equation of motion (EOM),
the parameter $\rho^\mathcal{P} = (m_u+m_q)/m_0^\mathcal{P}$ is introduced to reflect the quark masses terms in the EOM,
in our calacultion we only take into account the strange quark mass,
with neglecting the $u, d$ quark masses unless in the chiral masses $m_0^\mathcal{P}$.
To next-to-leading order in conformal spin and to the second moments in truncated conformal expansion of $\varphi_\mathcal{P}$, we get
\beq
\varphi_\mathcal{P}^P(x, \mu) &=& 1 + 3 \rho^\mathcal{P} \Big( 1-3a_1^\mathcal{P}+6a_2^\mathcal{P}\Big)(1+\ln x)
- \frac{\rho^\mathcal{P}}{2} \Big( 3-27a_1^\mathcal{P}+54a_2^\mathcal{P}\Big) \, C_1^{1/2}(2x-1) \non
&+& 3 \Big(10 \eta_{3\mathcal{P}} - \rho^\mathcal{P} (a_1^\mathcal{P} - 5 a_2^\mathcal{P}) \Big) \, C_2^{1/2}(2x-1)
+ \Big(10 \eta_{3\mathcal{P}} \lambda_{3\mathcal{P}} - \frac{9}{2}\rho^\mathcal{P}a_2^\mathcal{P} \Big) \, C_3^{1/2}(2x-1) \non
&-& 3 \eta_{3\mathcal{P}} \omega_{3\mathcal{P}} \, C_4^{1/2}(2x-1) \, ,
\label{eq:DA-t3-P}\\
\varphi_\mathcal{P}^\sigma(x, \mu) &=& 6x(1-x) \Big\{ 1 + \frac{\rho^\mathcal{P}}{2} \Big(2 - 15 a_1^\mathcal{P} + 30 a_2^\mathcal{P}\Big)
+ \rho^\mathcal{P} \Big(3a_1^\mathcal{P} - \frac{15}{2}a_2^\mathcal{P}\Big) \, C_1^{3/2}(2x-1) \non
&+& \frac{1}{2} \Big( \eta_{3\mathcal{P}}(10-\omega_{3\mathcal{P}}) + 3\rho^\mathcal{P} a_2^\mathcal{P} \Big) \, C_2^{3/2}(2x-1)
+ \eta_{3\mathcal{P}} \lambda_{3\mathcal{P}} \, C_3^{3/2}(2x-1) \non
&+& 3 \rho^\mathcal{P} \Big( 1- 3a_1^\mathcal{P} + 6 a_2^\mathcal{P} \Big) \, \ln x \Big\} \,,
\label{eq:DA-t3-T} \\
\varphi_{3\mathcal{P}}(x_i) &=& 360 x_1 x_2 x_3^2 \Big\{ 1 + \lambda_{3\mathcal{P}} (x_1-x_2) + \omega_{3\mathcal{P}} \frac{1}{2} (7x_3-3) \Big\} \,,
\label{eq:DA-t3-3p}
\eeq
where the contributions from the three-particle and from the two-particle by EOM are separated clearly,
the three parameters $f_{3\mathcal{P}}, \lambda_{3\mathcal{P}}, \omega_{3\mathcal{P}}$ can be defined by the matrix element of local twist-3 operators,
and their evolution have the mixing terms with the quark mass \cite{BallWN}.

For the two-particle twist-4 DAs, the definition considered in the strictly light-cone expansion in Eq.~(\ref{eq:DAs-A}) is more convenient to be used in the QCD calculation,
and their relations to the invariant amplitudes $\psi_{4\mathcal{P}}, \phi_{4\mathcal{P}}$ defined in the Lorentz structure are,
\beq
g_{2\mathcal{P}}(x) = -\frac{1}{2}\int_0^x dx' \psi_{4\mathcal{P}}(x') \,, \,\,\,\,\,
g_{1\mathcal{P}}(x) = \frac{1}{16} \phi_{4\mathcal{P}}(x) + \int_0^x dx' g_{2\mathcal{P}}(x') \, .
\label{eq:DA-t4-2p}
\eeq
The relations between different operators by EOM indicate that these Lorentz invariant amplitudes are written in terms of
the "genuine" twist-4 contribution from the three-particle DAs $\varphi_\parallel(x_i), \varphi_{\perp}(x_i)$
and the Wandzura-Wilczek-type mass corrections from the two-particle lower twist DAs,
distinguishing by parameters $\delta_{\mathcal{P}}^2$ and $m_\mathcal{P}^2$, respectively.
The corrected expressions are \cite{KhodjamirianYS}
\beq
\psi_{4\mathcal{P}}(x) &=& \delta_{\mathcal{P}}^2 \Big[ \frac{20}{3} \, C_2^{1/2}(2x-1) + \frac{49}{2} a_1^\mathcal{P} \, C_3^{1/2}(2x-1) \Big] \non
&+& m_\mathcal{P}^2 \Big\{
6 \rho^\mathcal{P} \Big( 1-3 a_1^\mathcal{P} +6a_2^\mathcal{P}\Big) \,C_0^{1/2}(2x-1) \non
&-& \Big[\frac{18}{5}a_1^\mathcal{P} + 3\rho^\mathcal{P}\Big(1-9a_1^\mathcal{P}+18a_2^\mathcal{P}\Big) + 12\kappa_{4\mathcal{P}}\Big] \, C_1^{1/2}(2x-1) \non
&+&\Big[ 2 - 6 \rho^\mathcal{P} \Big( a_1^\mathcal{P} - 5 a_2^\mathcal{P} \Big) + 60 \eta_{3\mathcal{P}} \Big] \, C_2^{1/2}(2x-1) \non
&+&\Big(\frac{18}{5}a_1^\mathcal{P} - 9 \rho^\mathcal{P} a_2^\mathcal{P} + \frac{16}{3}\kappa_{4\mathcal{P}} + 20 \eta_{3\mathcal{P}} \lambda_{3\mathcal{P}} \Big) \, C_3^{1/2}(2x-1) \non
&+&\Big(\frac{9}{4} a_2^\mathcal{P} - 6 \eta_{3\mathcal{P}} \omega_{3\mathcal{P}} \Big) \,C_4^{1/2}(2x-1) \Big\} \non
&+&6m_q^2 \Big(1-3a_1^\mathcal{P}+6a_2^\mathcal{P}\Big) \, \ln x \,,
\label{eq:DA-t4-t2-psi} \\
\phi_{4\mathcal{P}}(x) &=& \delta_\mathcal{P}^2 \Big\{ \Big( \frac{200}{3} + 196(2x-1)a_1^\mathcal{P} \Big) x^2\bar{x}^2 \non
&+& 21 \omega_{4\mathcal{P}} \Big( x\bar{x} (2+13x\bar{x}) + [2x^3(6x^2-15x+10)\ln x] + [x \leftrightarrow \bar{x}] \Big) \non
&-& 14 a_1^\mathcal{P} \Big(x\bar{x}(2x-1)(2-3x\bar{x}) - [2x^3(x-2) \ln x] + [x \leftrightarrow \bar{x}] \Big)
\Big\} \non
&+&m_\mathcal{P}^2 \Big\{
\frac{16}{3}\kappa_{4\mathcal{P}} \Big( x (2x-\bar{x}) (1-2x\bar{x}) + [5(x-2)x^3 \ln x] - [x \leftrightarrow \bar{x}] \Big) \non
&+& 4\eta_{3\mathcal{P}} x\bar{x} \Big[
60\bar{x} + 10 \lambda_{3\mathcal{P}} \Big((2x-1)(1-x\bar{x}) - (1-5x\bar{x}) \Big) \non
&-& \omega_{3\mathcal{P}} \Big( 3-21x\bar{x}+28x^2\bar{x}^2+3(2x-1)(1-7x\bar{x})\Big) \Big] \non
&-& \frac{36}{5}a_2^\mathcal{P} \Big(\frac{1}{4} x\bar{x}(4-9x\bar{x}+110x^2\bar{x}^2) + [x^3(10-15x+6x^2) \ln x] + [x \leftrightarrow \bar{x}] \Big) \non
&+& 4 x\bar{x} (1+3x\bar{x}) \Big( 1 + \frac{9}{5}(2x-1)a_1^\mathcal{P} \Big)  \Big\} \,,
\label{eq:DA-t4-t2-phi}
\eeq
with $\eta_{3\mathcal{P}}=f_{3\mathcal{P}}/(f_\mathcal{P}m_0^\mathcal{P})$.
It is noticed in Eq.~(\ref{eq:DA-t4-t2-psi}) that $\psi_{4\mathcal{P}}(x)$ has a logarithm end-point singularity for the finite quark mass,
while this singularity is not existed in $\phi_{4\mathcal{P}}(x)$.
The conformal expansion of three-particle twist-4 DAs reads:
\beq
\psi_\parallel(x_i) &=& 120x_1x_2x_3 \Big\{
\delta_{\mathcal{P}}^2 \Big[ \frac{21}{8}(x_1-x_2)\omega_{4\mathcal{P}} + \frac{7}{20}a_1^\mathcal{P}(1-3x_3) \Big] \non
&~& \hspace{1.5cm}  + m_\mathcal{P}^2 \Big[-\frac{9}{20}(x_1-x_2)a_2^\mathcal{P} + \frac{1}{3}\kappa_{4\mathcal{P}} \Big] \Big\}\,,
\label{eq:DA-t2-3p-A-para} \\
\psi_\perp(x_i) &=& 30x^2_3 \Big\{ \delta_\mathcal{P}^2 \Big[\frac{1}{3}(x_1-x_2)
+ \frac{7}{10}a_1^\mathcal{P} \Big( -x_3(1-x_3) + 3(x_1-x_2)^2\Big)
+ \frac{21}{4} \omega_{4\mathcal{P}}(x_1-x_2)(1-2x_3) \Big] \non
&~& \hspace{0.6cm}  + m_\mathcal{P}^2 (1-x_3) \Big[ \frac{9}{40}(x_1-x_2) - \frac{1}{3}\kappa_{4\mathcal{P}} \Big] \Big\}\,,
\label{eq:DA-t2-3p-A-perp} \\
\tilde{\psi}_\parallel(x_i) &=& -120x_1x_2x_3 \delta_\mathcal{P}^2 \Big\{
\frac{1}{3} + \frac{7}{4}a_1^\mathcal{P}(x_1-x_2) + \frac{21}{8}\omega_{4\mathcal{P}}(1-3x_3) \Big\} \,,
\label{eq:DA-t2-3p-V-perp} \\
\tilde{\psi}_\perp(x_i) &=& 30x_3^2 \Big\{
\delta_\mathcal{P}^2 \Big[ \frac{1}{3}(1-x_3) - \frac{7}{10}a_1^\mathcal{P}(x_1-x_2)(4x_3-3) +
\frac{21}{4}\omega_{4\mathcal{P}}(1-x_3)(1-2x_3) \Big] \non
&~& \hspace{0.6cm}  + m_\mathcal{P}^2 \Big[ \frac{9}{40}a_2^\mathcal{P} (x_1^2- 4x_1x_2+x_2^2) - \frac{1}{3}(x_1-x_2)\kappa_{4\mathcal{P}} \Big] \Big\} \,,
\label{eq:DA-t2-3p-V-perp}
\eeq
in which three nonperturbative parameters $\delta_{\mathcal{P}}^2, \omega_{4\mathcal{P}}, \kappa_{4\mathcal{P}}$ are introduced.
We close this section by noticing that all parameters in the conformal expansion of DAs have the scale dependence
and the behaviours of their evolutions can be found in Ref.~\cite{BallWN}.

\end{appendix}

}


\end{document}